\documentclass[sn-mathphys, icol, Numbered]{sn-jnl}

\usepackage{graphicx}%
\usepackage{multirow}%
\usepackage{amsmath,amssymb,amsfonts}%
\usepackage{amsthm}%
\usepackage{mathrsfs}%
\usepackage[title]{appendix}%
\usepackage{xcolor}%
\usepackage{textcomp}%
\usepackage{manyfoot}%
\usepackage{booktabs}%
\usepackage{algorithm}%
\usepackage{algorithmicx}%
\usepackage{algpseudocode}%
\usepackage{listings}%
\usepackage{natbib}
\usepackage{subcaption}

\begin{document}

\title[Detector development for the CRESST experiment]{Detector development for the CRESST experiment}


\author[1]{\fnm{G.} \sur{Angloher}}
\author[2,3]{\fnm{S.} \sur{Banik}}
\author[4]{\fnm{G.} \sur{Benato}}
\author[1,9]{\fnm{A.} \sur{Bento}}
\author[1]{\fnm{A.} \sur{Bertolini}}
\author[5]{\fnm{R.} \sur{Breier}}
\author[4]{\fnm{C.} \sur{Bucci}}
\author[2]{\fnm{J.} \sur{Burkhart}}
\author[1]{\fnm{L.} \sur{Canonica}}
\author[4]{\fnm{A.} \sur{D'Addabbo}}
\author[1]{\fnm{S.} \sur{Di Lorenzo}}
\author[2,3]{\fnm{L.} \sur{Einfalt}}
\author[6,10]{\fnm{A.} \sur{Erb}}
\author[6]{\fnm{F.v.} \sur{Feilitzsch}}
\author[2]{\fnm{S.} \sur{Fichtinger}}
\author[1]{\fnm{D.} \sur{Fuchs}}
\author[1]{\fnm{A.} \sur{Garai}}
\author[2]{\fnm{V.M.} \sur{Ghete}}
\author[4]{\fnm{P.} \sur{Gorla}}
\author[4]{\fnm{P.V.} \sur{Guillaumon}}
\author[2]{\fnm{S.} \sur{Gupta}}
\author[1]{\fnm{D.} \sur{Hauff}}
\author[5]{\fnm{M.} \sur{Ješkovsk\'y}}
\author[1,2]{\fnm{J.} \sur{Jochum}}
\author[6]{\fnm{M.} \sur{Kaznacheeva}}
\author[6]{\fnm{A.} \sur{Kinast}}
\author[2]{\fnm{H.} \sur{Kluck}}
\author[8]{\fnm{H.} \sur{Kraus}}
\author[7]{\fnm{S.} \sur{Kuckuk}}
\author[1]{\fnm{A.} \sur{Langenk\"amper}}
\author*[1]{\fnm{M.} \sur{Mancuso}}\email{mancuso@mpp.mpg.de}
\author[4,11]{\fnm{L.} \sur{Marini}}
\author[1]{\fnm{B.} \sur{Mauri}}
\author[7]{\fnm{L.} \sur{Meyer}}
\author[2]{\fnm{V.} \sur{Mokina}}
\author[4]{\fnm{M.} \sur{Olmi}}
\author[6]{\fnm{T.} \sur{Ortmann}}
\author[4,12]{\fnm{C.} \sur{Pagliarone}}
\author[4,6]{\fnm{L.} \sur{Pattavina}}
\author[1]{\fnm{F.} \sur{Petricca}}
\author[6]{\fnm{W.} \sur{Potzel}}
\author[5]{\fnm{P.} \sur{Povinec}}
\author[1]{\fnm{F.} \sur{Pr\"obst}}
\author*[1]{\fnm{F.} \sur{Pucci}}\email{frapucci@mpp.mpg.de}
\author[2,3]{\fnm{F.} \sur{Reindl}}
\author[6]{\fnm{J.} \sur{Rothe}}
\author[1]{\fnm{K.} \sur{Sch\"affner}}
\author[2,3]{\fnm{J.} \sur{Schieck}}
\author[6]{\fnm{S.} \sur{Sch\"onert}}
\author[2,3]{\fnm{C.} \sur{Schwertner}}
\author[1]{\fnm{M.} \sur{Stahlberg}}
\author[1]{\fnm{L.} \sur{Stodolsky}}
\author[7]{\fnm{C.} \sur{Strandhagen}}
\author[6]{\fnm{R.} \sur{Strauss}}
\author[7]{\fnm{I.} \sur{Usherov}}
\author[2]{\fnm{F.} \sur{Wagner}}
\author[6]{\fnm{M.} \sur{Willers}}
\author[1]{\fnm{V.} \sur{Zema}}

\affil[1]{\orgdiv{Max-Planck-Institut f\"ur Physik}, \orgaddress{ \postcode{D-80805}, \city{M\"unchen}, \country{Germany}}}
\affil[2]{\orgdiv{Institut f\"ur Hochenergiephysik der \"Osterreichischen Akademie der Wissenschaften},\orgaddress{ \postcode{A-1050}, \city{Wien}, \country{Austria}}}
\affil[3]{\orgdiv{Atominstitut, Technische Universit\"at Wien},\orgaddress{ \postcode{A-1020}, \city{Wien}, \country{Austria}}}
\affil[4]{\orgdiv{Laboratori Nazionali del Gran Sasso},\orgaddress{ \postcode{I-67100},\city{Assergi}, \country{Italy}}}
\affil[5]{\orgdiv{Comenius University, Faculty of Mathematics, Physics and Informatics},\orgaddress{ \postcode{84248}, \city{Bratislava}, \country{Slovakia}}}
\affil[6]{\orgdiv{Physik-Department, Technische Universit\"at M\"unchen},\orgaddress{ \postcode{D-85747}, \city{Garching}, \country{Germany}}}
\affil[7]{\orgdiv{Eberhard-Karls-Universit\"at T\"ubingen},\orgaddress{ \postcode{D-72076}, \city{T\"ubingen}, \country{Germany}}}
\affil[8]{\orgdiv{Department of Physics, University of Oxford},\orgaddress{ \postcode{D-72076}, \city{Oxford}, \country{United Kingdom}}}
\affil[9]{\orgdiv{also at: LIBPhys-UC, Departamento de Fisica, Universidade de Coimbra},\orgaddress{ \postcode{P3004 516}, \city{Coimbra}, \country{Portugal}}}
\affil[10]{\orgdiv{also at: Walther-Mei\ss ner-Institut f\"ur Tieftemperaturforschung},\orgaddress{ \postcode{D-85747}, \city{Garching}, \country{Germany}}}
\affil[11]{\orgdiv{also at: GSSI-Gran Sasso Science Institute},\orgaddress{ \postcode{I-67100}, \city{L'Aquila}, \country{Italy}}}
\affil[12]{\orgdiv{also at: Dipartimento di Ingegneria Civile e Meccanica, Università degli Studi di Cassino e del Lazio Meridionale},\orgaddress{ \postcode{II-03043}, \city{Cassino}, \country{Italy}}}
\affil[13]{\orgdiv{currently at: INFN, Sezione di Milano LASA},\orgaddress{ \postcode{I-20054}, \city{Milano}, \country{Italy}}}


\abstract{Recently low-mass dark matter direct searches have been hindered by a low energy background, drastically reducing the physics reach of the experiments.
In the CRESST-III experiment, this signal is characterised by a significant increase of events below 200 eV. As the origin of this background is still unknown, it became necessary to develop new detector designs to reach a better understanding of the observations. Within the CRESST collaboration, three new different detector layouts have been developed and they are presented in this contribution.}

\maketitle

\section{Introduction}\label{sec1}

CRESST is a direct detection dark matter experiment located in the underground laboratory of Gran Sasso (LNGS) \cite{Abdelhameed:2019hmk}. CRESST operates cryogenic calorimeters consisting of a scintillating target crystal instrumented with a Transition Edge Sensor (TES) coupled to an auxiliary cryogenic calorimeter to measure the scintillation light, called a Light Detector (LD) \cite{J.rothe}.  CRESST is one of the leading experiments in the search for dark matter particles with masses $\lesssim1$~GeV/c$^2$ (light dark matter) \cite{spindepCRESST}. \\
Recently, direct searches for light dark matter have been hindered by an unexplained background at low energy, which reduces the sensitivity of the experiments \cite{osti_1862233}. In the CRESST experiment, this background is characterised by a sudden increase in the number of events at energies below 200 eV \cite{LEEdescr}. The cause of this phenomenon is still unclear. The resulting energy spectrum is featureless and could be mistaken for a signal from dark matter interactions \cite{Billard_2022}. \\
Similar observations have been made in many other experiments that focus on low energies, such as \cite{CDMS, DAMIC, EDELWEISS, NEWSG, SENSEI, NUCLEUS, Ricochet, MINER}, and more. This observation has garnered the attention of many scientists and nowadays this background is well-known as the low energy excess (LEE) (see \cite{osti_1862233}).\\
Using the CRESST-III detector design, various tests and modifications to the detector design have been made in order to identify the source of the LEE, with the goal of obtaining as much information as possible while making minimal changes to the detector setup.

\section{Low energy excess hunting in CRESST}
\label{section2}
\begin{figure}[b]
\begin{center}
\includegraphics[width=0.6\linewidth, keepaspectratio]{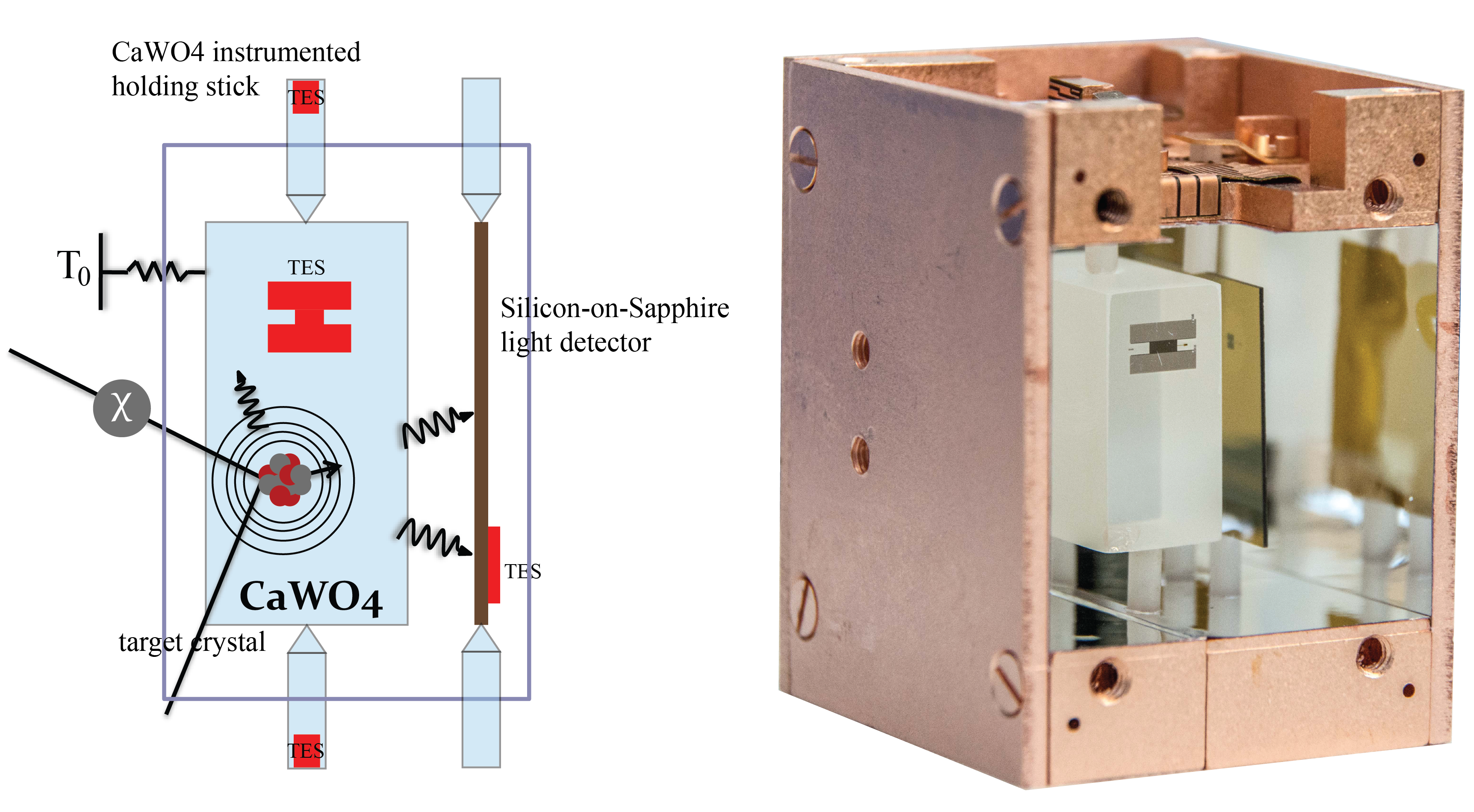}
\caption{{\it Left}: Schematic view of the detector design for CRESST-III modules. An energy deposition produces phonons, measured by the TES deposited on the target crystal, and photons, collected by the SOS light detector. Both detectors are weakly coupled to a thermal bath at T$_0\sim$10$\,$mK. A fully scintillating housing and instrumented CaWO$_4$ holder sticks are used to veto possible background originating from surrounding surfaces. {\it Right}: Picture of an open CRESST-III detector module.}
\label{scheme_det_A}
\end{center}
\end{figure}

The first observation of the LEE was made using a CRESST-III detector module in 2019 \cite{CRESST3det}. The CRESST-III detector module consists of a 20 $\times$ 20 $\times$ 10$\,$mm$^3$ CaWO$_{4}$ scintillating crystal weighing approximately $24\ g$ as main absorber and a smaller cryogenic calorimeter made of a 10 $\times$ 10 $\times$ 0.4$\,$mm$^3$ Silicon-On-Sapphire (SOS) as light detector. This detector concept is discussed in detail in \cite{CRESST3det}. 
Both calorimeters are equipped with a tungsten Transition Edge Sensors (W-TES) for temperature read out. The detectors are enclosed in a fully scintillating housing and equipped with instrumented holders (see figure \ref{scheme_det_A}).
\\During the first data taking campaign with these detectors, a rise in the event rate at threshold was observed by multiple detectors, but the energy spectra of the different modules were not compatible with each other, ruling out the possibility of a single common source for these events. 
Extensive studies were made to address some possible origins of the LEE and are summarised in the following:
\begin{enumerate}
    \item\textbf{Single photons hitting the TES sensor.} All non active scintillating parts inside the detector housing have been removed.
    \item\textbf{Internal relaxation of the lattice stress.} CaWO$_4$ crystals growth differently from different sources with measured internal stress have been measured and compared.
    \item\textbf{Events transmitted to the crystal due to holding structure.} The sticks made from CaWO$_4$ have been replaced with copper sticks or bronze clamps.
    \item\textbf{Material dependence.} Main absorbers of Al$_2$O$_3$, LiAlO$_2$ and Si have been tested.
    \item\textbf{Target geometry related.} Light detectors have been analysed as individual detectors to verify the impact of mass, surface and sensor sizes.
\end{enumerate}
Our studies allowed to exclude several hypotheses, bringing more insight to the understanding of the LEE, although its origin could not be pinpointed, yet.
Meanwhile, the rate of the LEE was observed to decay over time with two time-constants: a fast one (order of few days) and a long one (order of 200 days) \cite{LEEdescr}. The former was observed to reactivate following thermal cycles from a working temperature of few mK to a few tens of K. \\

\section{Next-generation detector}
In this section, we present three new detector modules, designed with the objective of understanding and vetoing LEE events while improving the sensitivity at low energy. 

\subsection{Mini-beaker module} 
\label{MiniB}
The first design aims at enabling the possibility to detect and tag surface radiation and events transmitted through the holder. Thus, this module features a $4 \pi$ veto and an instrumented holding structure.

\subsubsection{Description}
\begin{figure}[b]
\begin{center}
\includegraphics[width=0.4\linewidth, keepaspectratio]{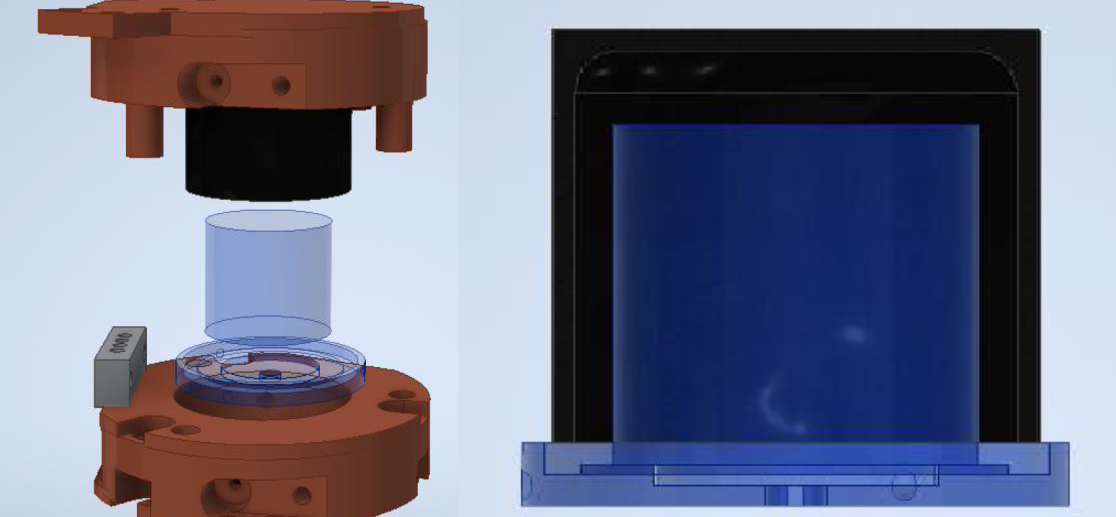}
\caption{CAD design of the mini-beaker module. The central cylindrical sapphire crystal is surrounded by auxiliary detectors to provide a 4$\pi$ veto. {\it Left}: Detailed CAD view of the various components of the module, including the bottom copper holder, the sapphire ring, the target crystal, the Si beaker, and the top copper holder. {\it Right}: Cross-sectional view of the three crystals in their working position. }\label{beaker-picture}
\end{center}
\end{figure}
The mini beaker module utilises a Al$_2$O$_3$ crystal with a cylindrical shape measuring 18$\,$mm in diameter and 15.7$\,$mm in height, resulting in a total volume of 4$\,$cm$^3$ equal to that of a standard CRESST-III target. The target is instrumented with a W-TES, directly deposited on a flat surface, and it is glued to a sapphire ring, on the same surface of the TES, through three spots of 50$\,$$\mu$m thick bi-component epoxy. The ring has a diameter of 25$\,$mm, is 3$\,$mm in height and it is instrumented with a W-TES, to allow for identification of events transmitted through the holding structure. The ring features a 3$\,$mm diameter opening in the centre to allow for electric contacts on the TES sensor of the target.
A silicon beaker, also instrumented with a W-TES, is surrounding the target, as illustrated in figure \ref{beaker-picture}. The beaker has an external diameter of 23$\,$mm and a height of 25$\,$mm. Together with the ring, the beaker provides a $4 \pi$ veto against radiation originating from the surrounding of the crystal. The ring and beaker are held in position with Cu-Be clamps. 
The complete module features 3 active channels: the sapphire cylindrical crystal is the main absorber, while the ring and the beaker serve as veto channels. 
To minimise thermal interference between the different components of the module, the heaters are directly deposited onto the W-TES. They are insulated from the thermometer through a 350$\,$nm thick layer of SiO$_2$. This approach results in improved operability and fewer constraints on the critical temperature of the sensors.\\

\subsection{cm-cube array module}
\label{cm3}
The second module is the cm-cube module. This design features a single array of four 1$\,$cm$^3$ absorber crystals and two light detectors, with a holding scheme designed to minimise the forces acting on the crystals. The aim of this holding scheme is to investigate the influence of forces acting on the crystals on the LEE.

\subsubsection{Description}
\label{cm3_D}
\begin{figure}[h]
\begin{center}
\includegraphics[width=0.4\linewidth, keepaspectratio]{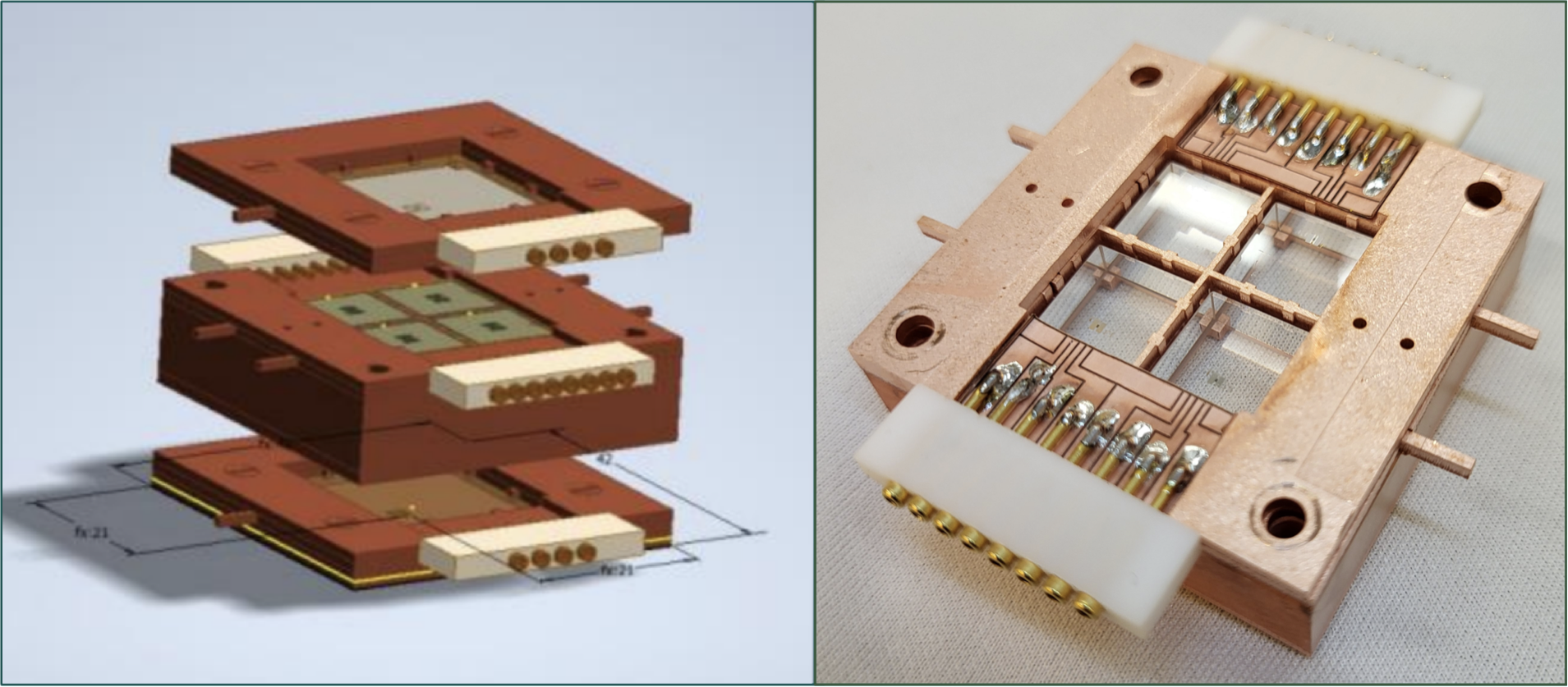}
\caption{Design of the cm-cube detector module. {\it Left}: Detailed CAD view of the various components of the module, the four main absorbers, the light detectors, and their copper holding structures. {\it Right}: A picture of the cm-cube main absorber crystals mounted in their test holder.}\label{cm-picture}
\end{center}
\end{figure}

The cm-cube module utilises four 1 $\times$ 1 $\times$ 1$\,$cm$^3$ CaWO$_4$ crystals, for the same total volume of 4$\,$cm$^3$ as the standard CRESST-III design. The module features two light detectors, one placed on top of the four absorber targets and one on the bottom. Thus, the complete module features six active channels. \\
The advantage of such a design on the one hand lies  in the possibility to veto coincident events and on the other hand, the smallness of the absorbers helps lowering the reachable threshold \cite{threshold}, a key parameter for the search of dark matter in CRESST. We expect that each of the cubes lowers the threshold by a factor of 2 compared to the standard CRESST-III target crystals. The novel holding scheme features targets placed on a minimal amount of copper and kept in place only by their weight, known as "gravity assisted" holder (see figure \ref{cm-picture}).

\subsection{DoubleTES module}
\label{doubleTES}
The final module we present is the doubleTES sensor module. The basic idea of this module is to equip the main target with two identical sensors, independently read out. This module allows to investigate the possibility that the origin of the low energy excess is sensor related.

\subsubsection{Description}
\begin{figure}[h]
\begin{center}
\includegraphics[width=0.9\linewidth, keepaspectratio]{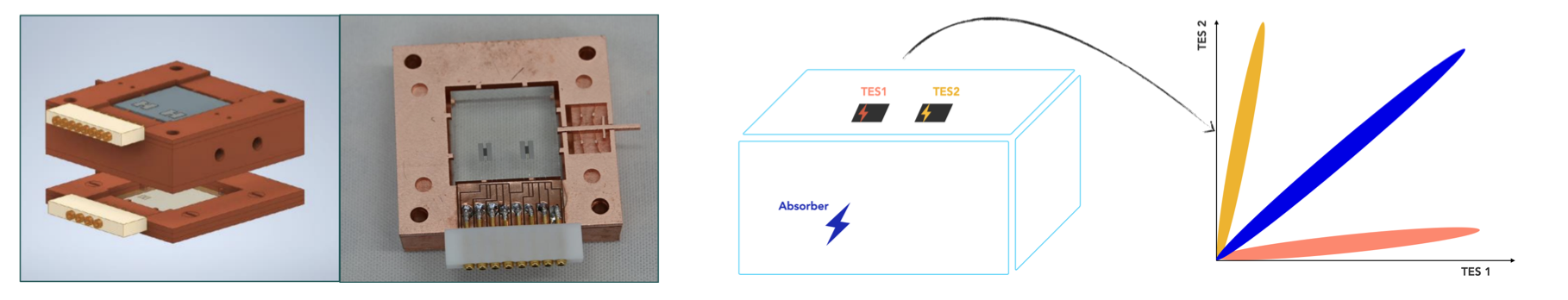}
    \caption{This figure illustrates the design of the doubleTES, one of the next-generation detector modules presented in this work. {\it Left}: Detailed view of the various components of the module. The holes on the side of the copper structure are used to easily place iron calibration sources when needed and an actual picture of the doubleTES crystal, showing the two identical TES seen from top.  {\it Right}: Schematic representation of the signal originating from different sources in the doubleTES. In a plot with the energies of the two TES on the axes, we expect a clear separation of the events originating from the crystal from those originating from the two different sensors.}\label{double-picture}
    \end{center}
\end{figure}

The doubleTES utilises the same crystals as the standard CRESST-III module: a CaWO$_4$ crystal as the main absorber and a SOS crystal as a light detector. The main absorber crystal is instrumented with two identical W-TES, directly deposited on the top of the crystal surface simultaneously. To minimise the thermal interference between the two sensors, the heaters are directly deposited onto the W-TES and insulated from the thermometer throught a 350$\,$nm thick layer of SiO$_2$, as for the mini beaker (see Section \ref{MiniB}). The light detector is instrumented with one W-TES only, also evaporated on the crystal surface, and it is placed under the main absorber as in figure \ref{double-picture} (left). 
The holding scheme utilised in this design is an adaptation of the "gravity assisted" holding scheme employed in the cm-cube module (see Section \ref{cm3}), aimed at minimising the forces acting on the crystal. \\
Combining the information obtained with the two sensors, it is possible to study the origin of the events (see figure \ref{double-picture} (right)). 

\section{Conclusions}
In summary, this contribution has introduced three cutting-edge detector modules designed for integration into the forthcoming CRESST-III measurement campaign, scheduled in early 2024. These innovative modules each play a pivotal role in advancing our pursuit of uncovering the origins of the LEE (Low-Energy Excess), promising exciting prospects for furthering our understanding of cryogenic detectors at low energy.


\bibliographystyle{sn-mathphys}
\bibliography{sn-bibliography}

\end{document}